\documentclass[twocolumn,pra,superscriptaddress,nofootinbib]{revtex4-2}
\pdfoutput=1
\usepackage[utf8]{inputenc}
\usepackage[english]{babel}
\usepackage[T1]{fontenc}
\usepackage[normalem]{ulem}
\usepackage{soul}

\usepackage{tikz}
\usepackage{lipsum}

\usepackage{graphicx,amsmath,color,bm,rotating,dsfont,stackrel,ulem,braket}
\usepackage[hyphenbreaks]{breakurl}
\usepackage[colorlinks=true,linkcolor=blue,citecolor=blue,urlcolor =blue]{hyperref}

\newcommand{\eq}[1]{\begin{equation}\begin{aligned}#1\end{aligned}\end{equation}}
\newcommand{\iu}{\text{i}}
\newcommand{\eu}{\text{e}}
\newcommand{\ha}{\hat{a}}
\newcommand{\had}{\hat{a}^\dagger\vphantom{a}}
\newcommand{\hb}{\hat{b}}
\newcommand{\hbd}{\hat{b}^\dagger\vphantom{a}}

\newcommand{\Tr}{\mathop{\mathrm{Tr}} \nolimits}

\newcommand{\Var}{\mathop{\mathrm{Var}}\nolimits}
\newcommand{\Cov}{\mathop{\mathrm{Cov}} \nolimits}
\newcommand{\vac}{\ket{\mathrm{vac}}}
\newcommand{\expct}[1]{\left\langle#1\right\rangle}
\newcommand{\sugg}[1]{{#1}}

\begin{document}
\setstcolor{red}

\title{Measuring the quadrature coherence scale on a cloud quantum computer}

\author{Aaron Z. Goldberg}
\affiliation{National Research Council of Canada, 100 Sussex Drive, Ottawa, Ontario K1N 5A2, Canada}
\affiliation{Department of Physics, University of Ottawa, Advanced Research Complex, 25 Templeton Street, Ottawa, Ontario K1N 6N5, Canada}

\author{Guillaume S. Thekkadath}
\affiliation{National Research Council of Canada, 100 Sussex Drive, Ottawa, Ontario K1N 5A2, Canada}

\author{Khabat Heshami}
\affiliation{National Research Council of Canada, 100 Sussex Drive, Ottawa, Ontario K1N 5A2, Canada}
\affiliation{Department of Physics, University of Ottawa, Advanced Research Complex, 25 Templeton Street, Ottawa, Ontario K1N 6N5, Canada}
\affiliation{Institute for Quantum Science and Technology, Department of Physics and Astronomy, University of Calgary, Alberta T2N 1N4, Canada}
  
\begin{abstract}
 Coherence underlies quantum phenomena, yet it is manifest in classical theories; delineating coherence's role is a fickle business. The quadrature coherence scale (QCS) was invented to remove such ambiguity, quantifying quantum features of any single-mode bosonic system without choosing a preferred orientation of phase space. The QCS is defined for any state, reducing to well-known quantities in appropriate limits including Gaussian and pure states, and, perhaps most importantly for a coherence measure, it is highly sensitive to decoherence. Until recently, it was unknown how to measure the QCS; we here report on an initial measurement of the QCS for squeezed light and thermal states of light. This is performed using Xanadu's machine Borealis, accessed through the cloud, which offers the configurable beam splitters and photon-number-resolving detectors essential to measuring the QCS. The data and theory match well, certifying the usefulness of interferometers and photon-counting devices in certifying quantumness.
\end{abstract}

\maketitle
\section{Introduction}
Coherence is essential to interference and enables quantum properties including superposition and entanglement \cite{MandelWolf1995,Glauber1963quantumtheorycoherence,Sudarshan1963,Kasevich2002,Leggett2002,Bloch2008,Streltsovetal2015}. There are many ways to quantify coherence \cite{Glauber1963quantumtheorycoherence,Sudarshan1963,Mandel1979,BachLuxmannEllinghaus1986,LeeJeong2011,FrowisDur2012,Lietal2012,Girolami2014,Monrasetal2014,Yuanetal2015,YadinVedral2016,Kwonetal2018,Zhangetal2018}, including its use as a resource \cite{Ekert1991,Scully1991,Aberg2006arxiv,Giovannettietal2011,Baumgratzetal2014,Lostaglioetal2015,Korzekwaetal2016,WinterYang2016,Streltsovetal2017,Messingeretal2020,GoldbergSteinberg2020,Fengetal2022arxiv}, which agree that coherence is related to the magnitudes of off-diagonal components of a density matrix.

Quantifying coherence is thus notoriously ambiguous because coherence is a basis-dependent quantity. 
This has long been a concern when trying to demonstrate macroscopic quantum effects \cite{Leggett1980,Sekatskietal2014,Frowisetal1015,Frowisetal2018}, from quantum effects in photosynthesis and light-harvesting compounds \cite{Engeletal2007,Panitchayangkoonetal2010,Lambertetal2013,WilkinsDattani2015,Brumer2018,Jianshuetal2020,Mancal2020,Calderonetal2022arxiv} to coherence in the human brain \cite{HameroffPenrose1996,Salarietal2011,KerskensLopezPerez2022} to entanglement of a tardigrade \cite{Vedral2021arxiv,Leeetal2022,Ross2022}.
One must always talk about coherence with respect to a particular basis, which can be tied to the eigenbasis of some particular observable.

The coherence scale associated with a particular operator $A$ can be defined for any state $\rho$ as the probability that $\rho$ couples two eigenstates of $A$, weighted by the distance between the respective two eigenvalues of $A$ \cite{HertzDeBievre2020}. In the case of a quantum system described by a single harmonic oscillator mode, which is ubiquitous in quantum optics and beyond, one can remove operator or basis ambiguity by defining the quadrature coherence scale (QCS) as the average of the coherence scales associated with the position and momentum operators \cite{deBievreetal2019,Horoshkoetal2019,Hertzetal2020,HertzDeBievre2020}. The QCS can be proven to be independent from the quadratures chosen, constitute a witness of quantumness for any state, and reduce to well-studied quantities for pure and Gaussian states with many desirable properties including providing a bound for the distance between a state and the set of classical states \cite{deBievreetal2019}. Until now, the QCS has not been measured; a recent proposal for measuring the QCS \sugg{with no reconfigurable parts} using a balanced beam splitter and photon-number-resolving detectors (PNRDs) changes that \cite{Griffetetal2022arxiv}.

We generate states and measure their QCSs using Xanadu's cloud quantum computer Borealis, which recently demonstrated quantum advantages in a Gaussian boson sampling task \cite{Madsenetal2022}. Borealis provides squeezed vacuum states for QCS measurements, as well as linear-optical networks capable of generating other states from squeezed vacuum states, such as thermal states, to have their QCSs measured. \sugg{Ideally, the measured QCSs will show squeezed states to have signatures of quantumness that thermal states do not possess. However,} because the networks are lossy and the detectors do not have perfect efficiency, we instead measure the QCS of lossy versions of squeezed vacuum and thermal states. Although we find that the QCS is not able to certify quantumness in squeezed states with this amount of loss, we demonstrate excellent \sugg{measurement} agreement with theoretically predicted QCS values for the variety of states tested here. We explain how to extrapolate these results to other settings and how to interpret a state for which the QCS alone cannot be used to certify quantumness; other forms of quantumness are still present in such states. As a byproduct of our protocol, we also use Borealis to directly measure purity and other properties of our quantum states. These together showcase the diversity of the QCS and the usefulness of Borealis and its underlying components.

\section{Mathematical Preliminaries}
Consider a single bosonic mode annihilated by $\ha$. Starting from the vacuum state, one can create Fock states with $n$ excitations via
\eq{
\ket{n}=\frac{\had^n}{\sqrt{n!}}\vac.
} In this context, the most classical states are agreed to be the canonical coherent states
\eq{
\ket{\alpha}=\eu^{-\tfrac{|\alpha|^2}{2}}\sum_{n=0}^\infty\frac{\alpha^n}{\sqrt{n}!}\ket{n}
} while the most quantum states differ from coherent states in some maximal way \cite{Goldbergetal2020extremal}. An arbitrary quantum state can be represented in the respective bases as
\eq{
\hat{\rho}=\sum_{m,n\geq 0}\rho_{mn}\ket{m}\bra{n}=\int 
 d^2\alpha P(\alpha)\ket{\alpha}\bra{\alpha}
} and must obey the positivity and trace constraints $\bra{\psi}\hat{\rho}\ket{\psi}\geq 0\,\forall\ket{\psi}$ and $\Tr(\hat{\rho})=1$. Many important notions of quantumness relate to the negativity or singularness of the $P$-function; a $P$-function that is positive and no more singular than a Dirac $\delta$ function indicates that a state is a convex combination of ``classical'' coherent states.

For such a single bosonic mode, the quadrature coherence scale (QCS) is defined as
\eq{
\mathcal{C}^2(\hat{\rho})=\frac{1}{2\mathcal{P}(\hat{\rho})}\left\{\Tr\left(\left[\hat{\rho},\hat{x}\right]\left[\hat{x},\hat{\rho}\right]\right)+\Tr\left(\left[\hat{\rho},\hat{p}\right]\left[\hat{p},\hat{\rho}\right]\right)\right\}
} for the purity $\mathcal{P}(\hat{\rho})=\Tr(\hat{\rho}^2)$ and the quadrature operators $\hat{x}=\left(\ha+\had\right)/\sqrt{2}$ and $\hat{p}=-\iu\left(\ha-\had\right)/\sqrt{2}$. This is the average of the coherence scales associated with the two quadrature operators and is equal to the same quantity with rotated quadratures $\hat{x}\to\hat{x}\cos\theta+\hat{p}\sin\theta$, $\hat{p}\to -\hat{x}\sin\theta+\hat{p}\cos\theta$. In the position and momentum representations formed from eigenstates $\ket{x}$ and $\ket{p}$ of the operators $\hat{x}$ and $\hat{p}$, respectively, the properties of the QCS are evident:
\eq{
\mathcal{C}^2(\hat{\rho})=\frac{1}{2}&\left[\int dx dx^\prime (x-x^\prime)^2\frac{\left|\bra{x}\hat{\rho}\ket{x^\prime}\right|^2}{
\int dy dy^\prime\left|\bra{y}\hat{\rho}\ket{y^\prime}\right|^2
}
\right.\\
&\left.
+\int dp dp^\prime (p-p^\prime)^2\frac{\left|\bra{p}\hat{\rho}\ket{p^\prime}\right|^2}{
\int dk dk^\prime\left|\bra{k}\hat{\rho}\ket{k^\prime}\right|^2
}\right].
} This quantifies the average of two quantities: the strength of the off-diagonal elements in the position basis, weighted by the difference between the positions in question, and the same for the momentum basis. The purity factors appearing in the denominators are necessary to ensure that the weights are normalized to represent a true probability distribution. As such, the QCS not only quantifies \textit{how much} coherence is present, it also quantifies \textit{where} the coherence is present, thereby giving more weight to macroscopic superpositions than their microscopic counterparts. Canonical coherent states have $\mathcal{C}^2=1$, from which one can infer that all states with classical $P$-functions have $\mathcal{C}^2\leq 1$. Conversely, larger values of the QCS signify the presence of quantumness, which is witnessed by
\eq{
\mathcal{C}^2>1.
}

For pure states, the QCS reduces to
\eq{
\mathcal{C}^2(\ket{\psi}\bra{\psi})=\Var_\psi(\hat{x})+\Var_\psi(\hat{p}),
} coinciding with well-studied indicators of quantumness including the total noise \cite{Schumaker1986,Hillery1989}, total variance \cite{Yadinetal2018}, mean quadrature variance \cite{Kwonetal2019}, and average quantum Fisher information for sensing displacements in phase space \cite{Goldbergetal2020extremal}. For Gaussian states centred at the origin of phase space, pure or mixed, the QCS takes the simple form
\eq{
\mathcal{C}^2(\hat{\rho}_{\mathrm{G}})=\left[\Var_{\hat{\rho}_{\mathrm{G}}}(\hat{x})+\Var_{\hat{\rho}_{\mathrm{G}}}(\hat{p})\right]\mathcal{P}(\hat{\rho}_{\mathrm{G}})^2,
} which is proportional to the trace of the inverse of the state's covariance matrix. These have been used to directly compare the quantum properties of a variety of states to each other and to establish the agreement of the QCS with other measures of quantumness in appropriate limits \cite{deBievreetal2019,HertzDeBievre2020,Hertzetal2020}.

The most quantum state we investigate here is squeezed vacuum:
\eq{
\ket{r\eu^{\iu\phi}}=\exp\left(-\frac{\had^2r\eu^{\iu\phi}-\ha^2r\eu^{-\iu\phi}}{2}\right)\vac.
} Such states are directly provided by Borealis and have QCS
\eq{
\mathcal{C}^2(r)=\cosh(2r),
}which is equivalent to $1+2\bar{n}$ with $\bar{n}=\expct{\had\ha}$, as for all pure states with $\expct{\ha}=0$. The quantumness grows with increasing squeezing magnitude $r$. 

The least quantum state we investigate here is a thermal state:
\eq{
\hat{\rho}_{\bar{n}}=\frac{1}{\bar{n}+1}\sum_{m=0}^\infty\left(\frac{\bar{n}}{\bar{n}+1}\right)^m\ket{m}\bra{m}.
} Such states are not directly provided by Borealis and have QCS
\eq{
\mathcal{C}^2(\hat{\rho}_{\bar{n}})=\frac{1}{1+2\bar{n}}.
}
The quantumness shrinks with increasing average number of excitations $\bar{n}$, corresponding to increasing temperature of the thermal state. To generate such states, we use the known result that a single mode of a two-mode squeezed vacuum state 
\eq{
\ket{\mathrm{TMSV}}_{ab}=\frac{1}{\sqrt{\cosh r}}\sum_{m=0}^\infty (\eu^{\iu\phi}\tanh{r})^m\ket{m}_a\otimes\ket{m}_b
}
is a thermal state
\eq{\Tr_b\left(\ket{\mathrm{TMSV}}_{ab}\bra{\mathrm{TMSV}}\right)=\frac{1}{{\cosh r}}\sum_{m=0}^\infty \tanh^{2m}{r}\ket{m}_a\bra{m}}
along with the remarkable property that a careful phase relationship between two single-mode squeezed states impinging on a symmetric beam splitter will generate a two-mode squeezed-vacuum state:\eq{
\hat{\mathrm{BS_S}}\ket{r\eu^{\iu\phi+\iu\tfrac{\pi}{2}}}_a\otimes\ket{r\eu^{\iu\phi+\iu\tfrac{\pi}{2}}}_b=\ket{\mathrm{TMSV}}_{ab},
} where the second mode is annihilated by $\hb$ and the beam splitter enacts
\eq{
\hat{\mathrm{BS_S}}\begin{pmatrix}
\ha\\
\hb
\end{pmatrix}\hat{\mathrm{BS_S}}^\dagger=\frac{1}{\sqrt{2}}\begin{pmatrix}
1&\iu\\\iu&1
\end{pmatrix}\begin{pmatrix}
\ha\\
\hb
\end{pmatrix}.
} The latter relationship is at the heart of the entanglement-generating property of beam splitters, where a different phase relationship will leave the two inputs unchanged and thereby generate zero entanglement \cite{HuangAgarwal1994,KimSanders1996,Kimetal2002,Xiangbin2002,Wolfetal2003,Jiangetal2013,GoldbergJames2018,GoldbergHeshami2021}.

True experiments exist in the laboratory \cite{Peres1997}, so the quantum states that will be measured are the above states subject to loss, detector inefficiencies, and other imperfections. Combining the majority of these effects into a single aggregate transmission probability $\eta$, without loss of generality, we can consider the loss transformation
\eq{
\ha\to\sqrt{\eta}\ha+\sqrt{1-\eta}\hat{v}
} to have acted on the initial states and take all expectation values assuming the orthogonal vacuum mode annihilated by $\hat{v}$ to initially be unpopulated. This yields the QCS for the lossy squeezed-vacuum state [see Appendix \ref{app:lossy squeezed QCS} Eq. \eqref{eq:QCS lossy Gaussian}] \eq{
\mathcal{C}^2(r;\eta)=\frac{1}{1+(1-2\eta)\frac{\eta\cosh(2r)-\eta}{\eta\cosh(2r)-\eta+1}},
} which is greater than one for any nonzero $r$ and $\eta>1/2$; with $\eta>1/2$, the QCS of squeezed vacuum increases monotonically with $r$. For the lossy thermal states, which transform into other thermal states with diminished energies  $\bar{n}\to\eta\bar{n}$, the QCS becomes \eq{\mathcal{C}^2(\hat{\rho}_{\bar{n}};\eta)=\frac{1}{1+2 \eta\bar{n} }.
\label{eq:QCS Gaussian}
} Since these are both Gaussian states, the QCS diminishes due to $\left[\Var_{\hat{\rho}_{\mathrm{G}}}(\hat{x})+\Var_{\hat{\rho}_{\mathrm{G}}}(\hat{p})\right]\to\eta \left[\Var_{\hat{\rho}_{\mathrm{G}}}(\hat{x})+\Var_{\hat{\rho}_{\mathrm{G}}}(\hat{p})-1\right]+1$, with ``interesting'' nonmonotonic dependence on $\eta$ coming from the transformation of the purity $\mathcal{P}$ when a state is subject to loss (again, see Appendix \ref{app:lossy squeezed QCS} for an explicit form\sugg{, which has not appeared before to the best of our knowledge}). In the appropriate limit of complete loss, every state becomes a zero-amplitude coherent state with $\mathcal{C}^2=1$.

\section{Setup}
To measure the QCS of a particular state, one requires access to two copies of that state, a balanced beam splitter, and photon-number-resolving detectors (PNRDs). This can be seen by expressing the QCS as \cite{Griffetetal2022arxiv}
\eq{
\mathcal{C}^2(\hat{\rho})=
\sugg{
\frac{\Tr\left[\hat{\mathrm{BS_B}}\left(\hat{\rho}\otimes \hat{\rho}\right)\hat{\mathrm{BS_B}}^\dagger (-1)^{\had\ha}(1+2\had\ha)\right]}
{\Tr\left[\hat{\mathrm{BS_B}}\left(\hat{\rho}\otimes \hat{\rho}\right)\hat{\mathrm{BS_B}}^\dagger (-1)^{\had\ha}\right]}
},
} where the balanced beam splitter enacts
\eq{
\hat{\mathrm{BS_B}}\begin{pmatrix}
\ha\\
\hb
\end{pmatrix}\hat{\mathrm{BS_B}}^\dagger=\frac{1}{\sqrt{2}}\begin{pmatrix}
1&1\\ 1&-1
\end{pmatrix}\begin{pmatrix}
\ha\\
\hb
\end{pmatrix}.
} Each component of this expression is provided by Borealis; technical specifications and diagrams of the setup can be found in Ref. \cite{Madsenetal2022}. \sugg{The code used to deploy these circuits on Borealis is available in our Github repository at \url{https://github.com/AaronGoldberg9/QCS_Borealis_public}. Importantly, this same setup can be used to measure the QCS of \textit{any} input state without changing any measurement configurations.}

\begin{figure*}
    \centering
    \includegraphics[width=\textwidth,trim={2cm 4cm 2cm 4cm},clip]{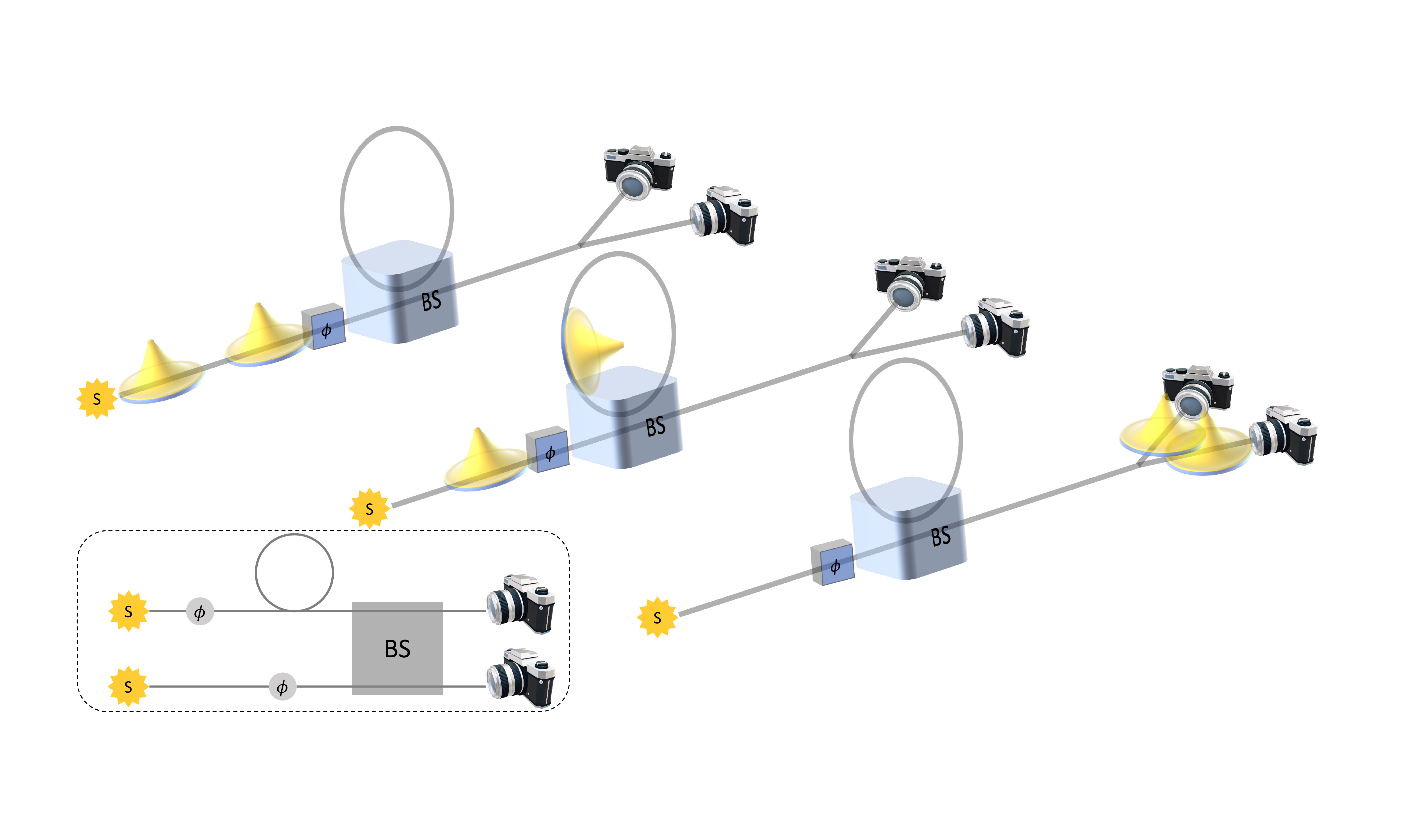}
    \caption{Schematic for measuring the QCS of a squeezed-vacuum state of light. Two copies of the same squeezed-vacuum state are created by the source $S$, one gets stored in a delay line \sugg{equal to the pulse separation of $\tau=1/(6\,\mathrm{MHz})$}, then the second gets sent through a phase shift $\phi$ before impinging upon a symmetric beam splitter $\mathrm{BS}$ coincidentally with the first photon. The phases are chosen so that one mode destructively interferes, the photon numbers are measured for each mode by a PNRD, then the photon-number distribution from the mode with destructive interference is used to infer the QCS. Loss and inefficiencies are present throughout and can be collected as one large loss channel acting identically on each input state. Inset: the schematic ``unravelled'' to represent the time-bin modes as spatial modes.}
    \label{fig:schematic SV}
\end{figure*}

\begin{figure*}
    \centering
    \includegraphics[width=\textwidth,trim={1.5cm 1cm 1.5cm 1cm},clip]{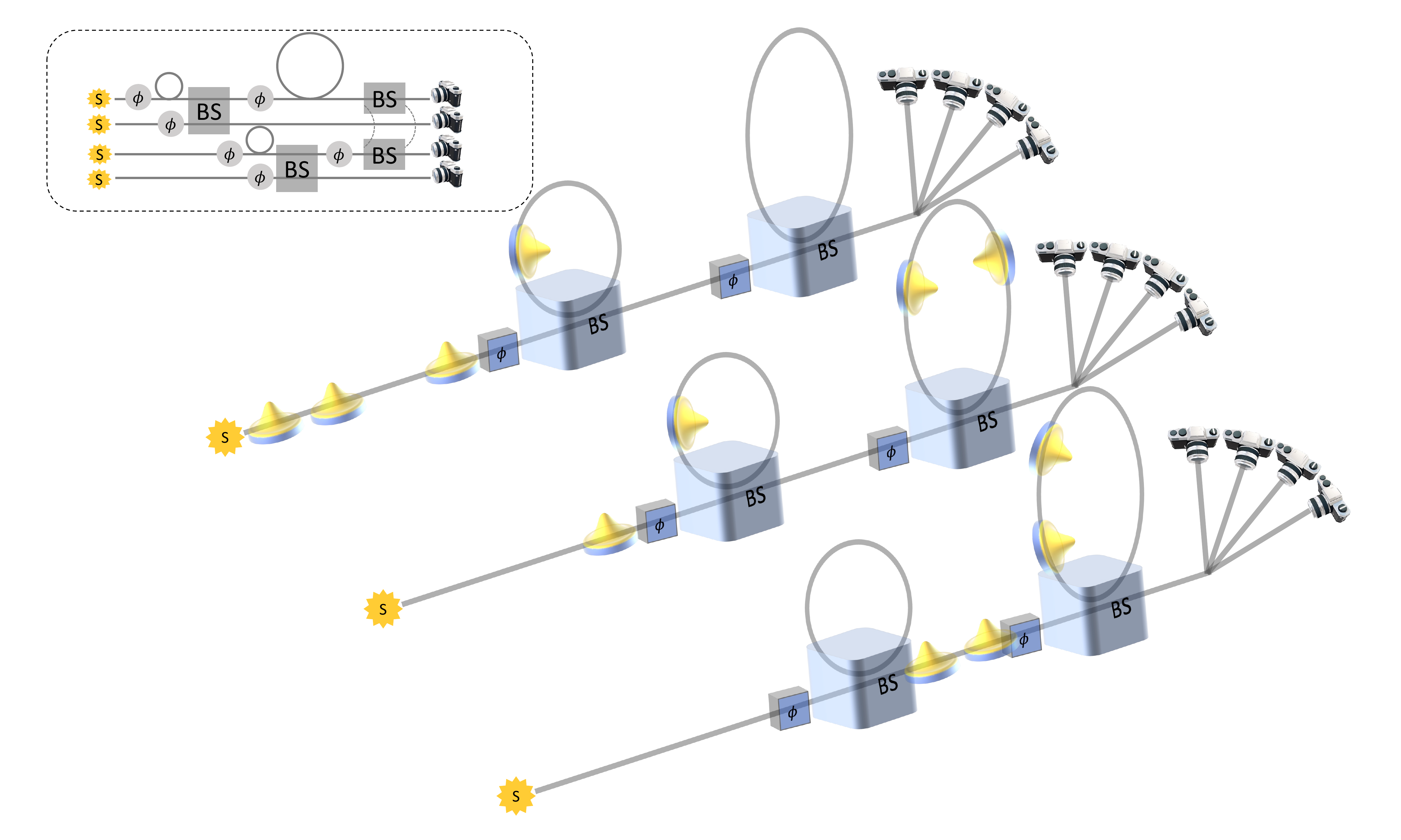}
    \caption{Schematic for measuring the QCS of a thermal state of light. Two copies of the same squeezed-vacuum state are created by the source $S$ and then interfere as in Fig. \ref{fig:schematic SV}, but with a different phase $\phi$ to generate a TMSV state. Both branches of the TMSV are sent into a second, larger delay loop \sugg{with period $6\tau$} while the first delay loop is reused to generate another copy of the TMSV from another two squeezed-vacuum states. The time bins are chosen such that the first branch of the first TMSV exits the second delay loop coincident with the first branch of the second TMSV arriving there. The phases are now chosen so that one mode destructively interferes, the photon numbers are measured for each mode by a PNRD, then the photon-number distribution from the mode with destructive interference is used to infer the QCS. Loss and inefficiencies are present throughout and can be collected as one large loss channel acting identically on each input state. Inset: the schematic ``unravelled'' to represent the time-bin modes as spatial modes.}
    \label{fig:schematic Th}
\end{figure*}

First, multiple copies of identical squeezed vacuum states are created by pumping an optical parametric oscillator repeatedly with identical pulses carved from a single continuous-wave laser source \sugg{upconverted to 775 nm pulses repeated at 6 MHz with average power 3.7 mW and duration of 3 ns per pulse}. This prepares a state $\ket{r\eu^{\iu\phi}}^{\otimes m}$ for a desired number of time-bin modes $m$ \sugg{each separated by $\tau=1/(6\, \mathrm{MHz})$}, where $r$ is chosen to be either $0.653$, $0.978$, or $1.156$. The pulse trains are sent to a series of delay loops of different lengths with configurable phase shifters and beam splitters at the entrance of each loop. To interfere a particular pair of time-bin modes, the earlier time-bin mode must be sent through a delay loop to exit coincidentally with the later time-bin mode arriving at the loop. Then, after all of the interference occurs, the time-bin modes are each sent to one of 16 transition-edge sensors that measures the total number of photons in the corresponding mode. \sugg{Calibration data for the efficiencies and phases imparted by each of these components are recorded in our Github repository.} The photon-number statistics of the final global state $\hat{\varrho}$
\eq{
p(n_a,\cdots,n_m)=\bra{n_a}_a\otimes\cdots\otimes\bra{n_m}_m\hat{\varrho}
\ket{n_a}_a\otimes\cdots\otimes\ket{n_m}_m
} can then be used to determine the QCS and other relevant properties of the input state $\hat{\rho}$. These are determined by performing between $N=9.5\times 10^5$ and $10^6$ trials and counting the number of times each arrangement of photon numbers is recorded. A figure of the general setup \sugg{as well as more details of the squeezed light generation, the beam splitters, the delay loops, and the detectors} can be found in Ref. \cite{Madsenetal2022}; we sketch our specific setups in Figs. \ref{fig:schematic SV} and \ref{fig:schematic Th}.

\sugg{Any} experiment to measure the QCS for squeezed vacuum requires two squeezed vacuum states; $m=2$. As depicted in Fig. \ref{fig:schematic SV}, the first mode is sent through the first delay line, interacts with the second mode on a balanced beam splitter after the latter arrives at the delay and the former traverses the delay, then the number of photons in each mode is measured. Ignoring the results of the measurement in mode $a$ [i.e., computing the marginal $p_{n_b}=\sum_{n_a,n_c,\cdots,n_m}p(n_a,n_c,\cdots,n_m)$] allows us to compute the QCS as \cite{Griffetetal2022arxiv}
\eq{
\mathcal{C}^2(\hat{\rho})=1+2\frac{\sum_{n_b}n_b(-1)^{n_b}p_{n_b}}{\sum_{n_b}(-1)^{n_b}p_{n_b}},
\label{eq:QCS from PNRD}
} where one may notice that the expression in the denominator is the \sugg{\textit{parity}} of the output state, exactly equal to the \sugg{\textit{purity}} of the input state: $\mathcal{P}=\sum_{n_b}(-1)^{n_b}p_{n_b}$. We can then assign errors to this value using multinomial statistics.

Measuring the QCS for thermal light requires extra steps. As depicted in Fig. \ref{fig:schematic Th}, four identical squeezed-vacuum states are created, the first from each pair traverses the first delay loop to then interfere with the second from each pair, and the interference creates a TMSV in each pair with $2\sinh^2 r$ photons on average. The reduced state for each of the four modes is a thermal state with 
 an average of $\bar{n}=\sinh^2 r$ photons. Once these identical thermal states are created, we can determine the QCS from any two of them. We send the first time-bin mode to wait at the second delay loop until the first of the second pair of time-bin modes arrives. These then interact at a balanced beam splitter, the photon-number statistics of the modes are measured, and the marginal statistics in the mode with destructive interference are used to compute the QCS.

The Borealis device supplies calibration data from which one can compute the total transmission parameter $\eta$. When the loss is not identical among the time-bin modes, we can pretend to add loss throughout the circuit until the loss is equal throughout, without ``loss'' of generality.
This means that the overall transmission $\eta$ must be the lowest of all the relevant time-bin modes. We can also find an upper bound for $\eta$ by dividing the average number of photons arriving at a detector by the input energy $\sinh^2 r$. There will be more uncertainty regarding $\eta$ as one increases the time between when the device was calibrated and when the \sugg{demonstration} is performed, so we use both the quoted values of $\eta$ and the energy calculation for inspecting our \sugg{measurement} results. When demonstrating quantum computational advantage, the Borealis machine had at most $\eta\leq 0.482\pm0.009$ in any mode \cite{Madsenetal2022}, so we do not expect to certify quantumness in this \sugg{demonstration} due to the requirement $\eta>1/2$ for lossy squeezed states to have $\mathcal{C}^2(r,\eta)>1$.

Finally, we must model the uncertainty in our determination of the QCS, requiring an estimate for the uncertainty on each \textit{probability} that enters into Eq. \eqref{eq:QCS from PNRD}. 
For a single PNRD, each trial measures a particular number of photons. Assuming some arbitrarily large number of photons to be the maximum that can be measured, the data from these trials are multinomially distributed, with some underlying probability to observe $n$ photons. We estimate that underlying probability using the observed probability distribution by counting the total number of times $N_n$ that $n$ photons are observed: $p_n=N_n/N$. The estimated sample covariances of these estimates are given by 
$\Cov(p_m ,p_n )=\tfrac{1}{N-1}p_n(\delta_{mn}-p_m)$.
Differentiating the QCS with respect to the underlying probabilities yields
\eq{
\Var(\mathcal{C}^2)&=\sum_{mn}\Cov(p_m ,p_n )\frac{\partial \mathcal{C}^2}{\partial p_m}\frac{\partial \mathcal{C}^2}{\partial p_n}\\
&=\sum_{mn}\frac{p_n(\delta_{mn}-p_m)}{N-1}\frac{(-1)^{m+n}}{\mathcal{P}^2}(2m+1-\mathcal{C}^2)\\
&\quad\times(2n+1-\mathcal{C}^2)\\
&=\frac{1}{\mathcal{P}^2}\sum_{n}\frac{p_n}{N-1}\left(2n+1-\mathcal{C}^2\right)^2,
\label{eq:uncertainty QCS}
} where each component is evaluated at the observed values of $p_n$ and the final expression can be recast in terms of the measured purity, QCS, and moments of the measured mode's photon number $\had_{\mathrm{out}}\ha_{\mathrm{out}}$. The multinomial statistics reduce to Poisson statistics because $\sum_m p_m (-1)^m (2m+1-\mathcal{C}^2)=0$.

\section{Results}
\begin{figure}
    \centering\includegraphics[width=\columnwidth]{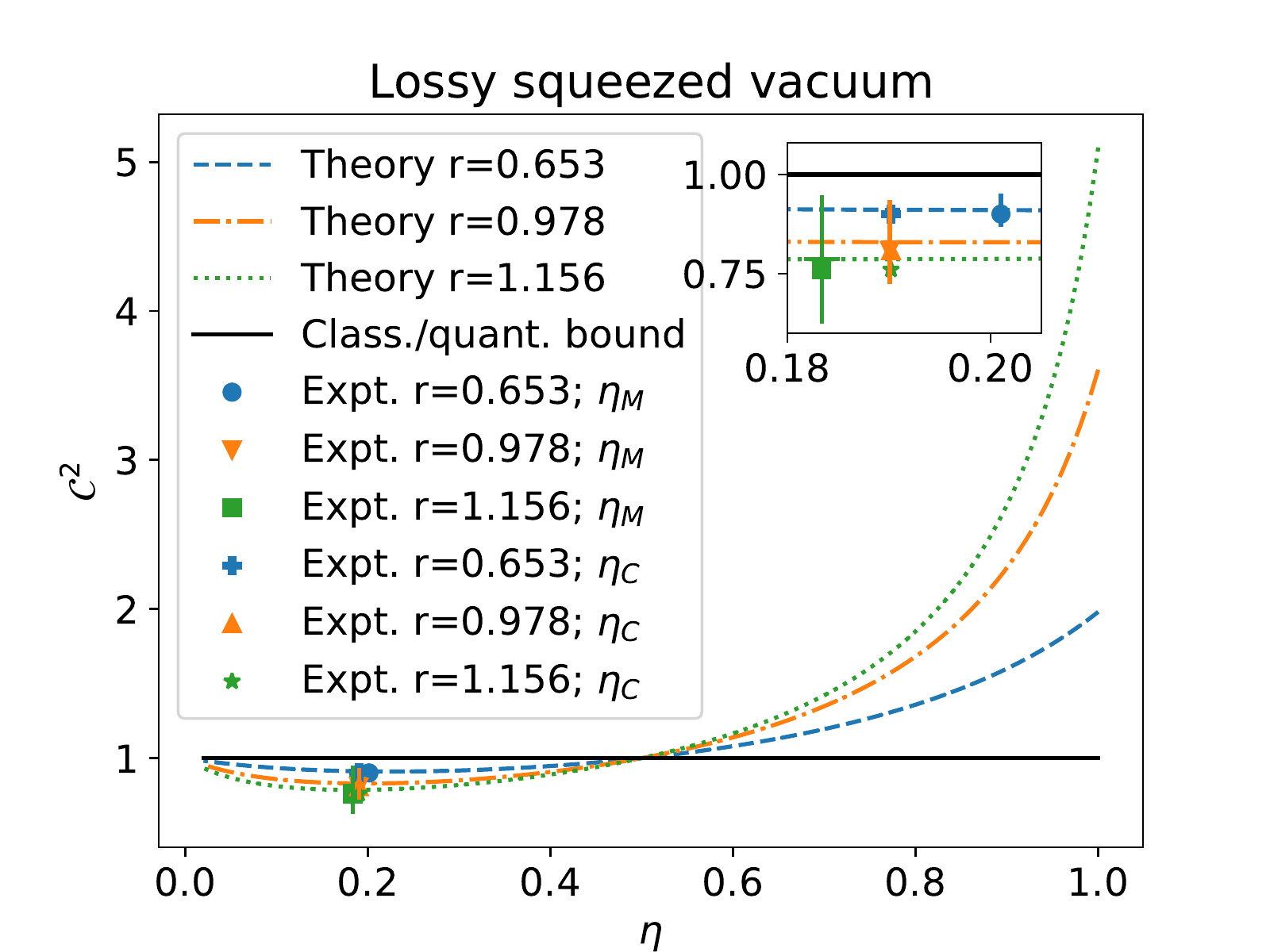}
    \caption{Expected QCS for a single-mode squeezed vacuum with squeezing parameter $r$ (energy $\sinh^2 r$) that has been subject to loss (overall transmission $\eta$); all parameters are unitless. Includes three data points from the Borealis machine. \sugg{Since $\eta$ is in principle not known \textit{a priori}, } the $\eta$ value \sugg{ascribed to the measured data point} is plotted at the calibrated point \sugg{supplied by a query to the Borealis machine} ($\eta_C$) and the point measured using the transmitted energy ($\eta_M$). The zoomed-in inset shows that the measured QCS is always slightly smaller than the predicted one. Errors for the data are too small to see; errors for the theory curves are given by ascribing uncertainties to the measured photon-number distribution. QCS less than or equal to unity cannot be used to certify quantumness, even if the latter is present in some other form.}
    \label{fig:QCS vs loss SV}
\end{figure}

\begin{figure}
    \centering\includegraphics[width=\columnwidth]{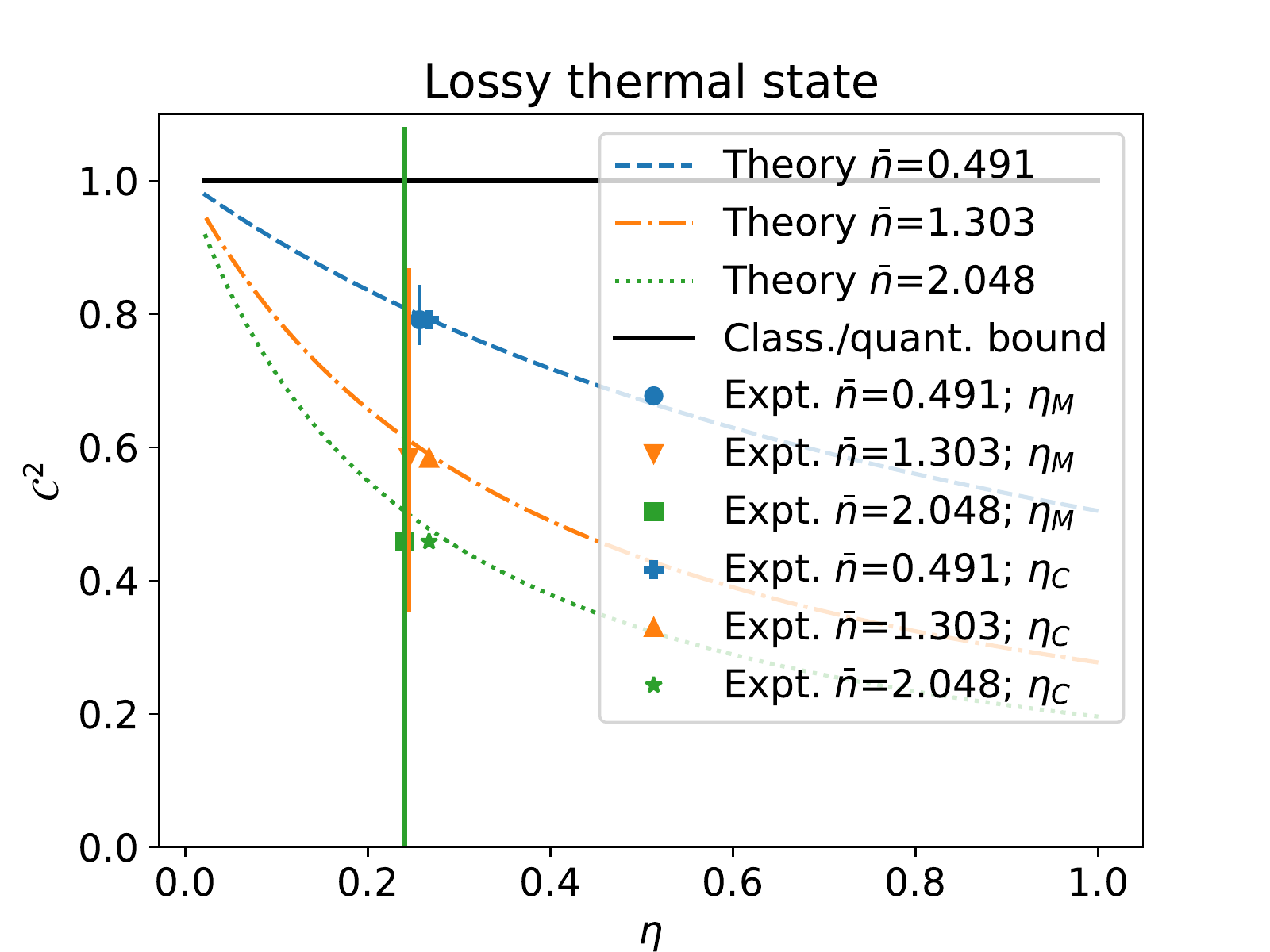}
    \caption{Expected QCS for a thermal state with average energy $\bar{n}$ that has been subject to loss (overall transmission $\eta$); all parameters are unitless.  Includes three data points from the Borealis machine. \sugg{Since $\eta$ is in principle not known \textit{a priori}, } the $\eta$ value \sugg{ascribed to the measured data point} is plotted at the calibrated point \sugg{supplied by a query to the Borealis machine} ($\eta_C$) and the point measured using the transmitted energy ($\eta_M$). The measured QCS is always slightly smaller than the predicted one. Errors for the data are too small to see; errors for the theory curves are given by ascribing uncertainties to the measured photon-number distribution.}
    \label{fig:QCS vs loss Th}
\end{figure}

The results are given in Table \ref{tab:res new} and depicted as points in Figs. \ref{fig:QCS vs loss SV} and \ref{fig:QCS vs loss Th}\sugg{; all of the raw data and the scripts used for processing them are available in our Github repository}. The uncertainties in the QCS computed using Eq. \eqref{eq:uncertainty QCS} are too small to be seen. The data always yield a slightly smaller QCS than predicted, possibly due to other \sugg{realistic} imperfections such as decoherence of the squeezed vacuum states that uniformly reduces the QCS and spurious detector counts that cannot be absorbed into the model for $\eta$. 

Spurious count rates are likely on the order of $10^{-3}$ \cite{Madsenetal2022}, including light leakage other than detector dark counts \cite{GoldbergHeshami2022xanadu}, for example from mode-mismatched photons \cite{Shchesnovich2014}, which are not enough to explain the discrepancy between the theory and \sugg{observations}. Instead, we look for errors on the theory curve due to discrepancies between the measured distribution $\{p_n\}$ and the expected distribution from a lossy squeezed state $\{\tilde{p}_n(r,\eta)\}$. Writing $\Cov(p_m,p_n)=\delta_{mn}(p_n-\tilde{p}_n)^2/3$ to consider a flat distribution of half-width $|p_n-\tilde{p}_n|$, we can compute the variance as per Eq. \eqref{eq:uncertainty QCS} for the theoretical result and plot this as the error bar on the theory curves in Figs. \ref{fig:QCS vs loss SV} and \ref{fig:QCS vs loss Th} centred at the value of $\eta_M$ for each measurement. The errors are larger for larger squeezed states, which are subject to greater decoherence, the errors are larger for thermal states than squeezed vacuum, which is likely due to the extra steps required to generate the thermal light, and the data points fall well within these ranges of error bars.
The amount of loss is sufficiently high that the QCS cannot certify quantumness for the lossy squeezed-vacuum states, always finding $\mathcal{C}^2<1$, but the close agreement between the data and theory for squeezed states implies that the QCS could readily certify quantumness if less loss was present. 

The measured transmission parameters listed in Table \ref{tab:res new} and depicted in Figs. \ref{fig:QCS vs loss SV} and \ref{fig:QCS vs loss Th} diminish with increasing probe state energy. This could be due to greater mode mismatch for the beams of light with more light or due to deviations from the calibrated squeezing strengths that are more pronounced at higher squeezing. 

Even though the states measured here have $\eta<1/2$ and so cannot be certified quantum by the QCS, they still have other quantum properties. It turns out that such states have negative $P$-functions for any transmission value $\eta>0$, as can be seen by performing a scaling transformation $P(\alpha)\to P(\alpha/\sqrt{\eta})$ when any state undergoes loss,\footnote{This can also be demonstrated by showing that any lossy squeezed-vacuum state will generate entanglement when impinging on a beam splitter with a vacuum state in the other input port, a necessary and sufficient condition for $P$-nonclassicality \cite{Asbothetal2005}, by using known conditions for bipartite entanglement of Gaussian states.
It also accords with the bound from Ref. \cite{Hertzetal2020} that says entanglement negativity can only be nonzero if $\mathcal{C}^2>\exp(-2/\eu)\approx 0.48$, which is the case here with minimum of 1/2 at $r\to\infty$.
} so such states can still be useful for some quantum information protocols. For example, Gaussian boson sampling can still have quantum advantages with $\eta\approx 0.3$ \cite{Madsenetal2022}. In contrast, QCS less than unity implies no possible quantum advantage in displacement sensing with the present state. This means that we have demonstrated the usefulness of this device for measuring the QCS, but that the quantum states it provides are insufficient so as to be certified quantum by the QCS and, therefore, that the quantum states measured here may only be useful for specific quantum applications.

\begin{table}[]
\centering
\caption{ Measured QCS values for squeezed-vacuum and thermal states of light with average energy: $\bar{n}=\sinh^2 r$. The transmission parameter for the initial state calculated using the devices calibration parameters is $\eta_{\mathrm{C}}$, while that from the measurement of the transmitted energy is $\eta_{\mathrm{M}}$. The transmission parameters are different for different \sugg{scenarios} because (a) the \sugg{demonstrations} are not conducted simultaneously and (b) the squeezed light has to traverse extra delay loops relative to the thermal light for technical reasons. The numbers in parentheses are the square roots of the calculated variances from the observed data and multinomial statistics.}
\label{tab:res new}
\begin{tabular}{ccccccc}
                              & \multicolumn{3}{c}{Squeezed vacuum}      & \multicolumn{3}{c}{Thermal light} \\
                              & QCS     & $\eta_{\mathrm{C}}$ & \multicolumn{1}{c|}{$\eta_{\mathrm{M}}$} & QCS           & $\eta_{\mathrm{C}}$     & $\eta_{\mathrm{M}}$     \\ \cline{2-7} 
\multicolumn{1}{c|}{$r=0.653$}  & $0.9003(9)$ & 0.190   & \multicolumn{1}{c|}{0.2010(2)}   & $0.792(1)$       & 0.267       & 0.2564(2)       \\
\multicolumn{1}{c|}{$r=0.978$}  & $0.809(2)$ & 0.190   & \multicolumn{1}{c|}{0.1901(8)}   & $0.584(3)$      & 0.267       & 0.2447(8)       \\
\multicolumn{1}{c|}{$r=1.156$} & $0.760(3)$ & 0.190   & \multicolumn{1}{c|}{0.183(2)}   & $0.459(5)$       & 0.267       & 0.240(2)      
\end{tabular}
\end{table}


\section{Discussion and conclusions}
The QCS is an intriguing quantifier of the amount of quantumness in any single-mode bosonic state. We measured the QCS for squeezed vacuum and thermal light with various average energies using the squeezed light, programmable interferometer, and PNRDs provided by Xanadu's Borealis. We found a maximal QCS of \sugg{$0.9003(9)$ and $0.792(1)$} for the squeezed and thermal states, respectively, which occured for the smallest input energies $\bar{n}\approx 0.5$ that we tested due to the amount of loss present in the system. The distance between any state and the set of $P$-classical states is lower bounded by $\mathcal{C}-1$ \cite{deBievreetal2019}, so QCS values smaller than $1$ cannot distinguish between $P$-classical and $P$-nonclassical states.
The amount of loss present prohibited certification of quantumness using the QCS; with transmission probability $\eta>1/2$, larger energies would have led to larger witnesses of quantumness for the squeezed-vacuum states \sugg{to distinguish them from the thermal states}. Since the data closely followed the theory, it follows that the present scheme, developed in Ref. \cite{Griffetetal2022arxiv}, could certainly be used to certify quantumness using PNRDs to measure the QCS if less loss was present. 

Most of the trials reported on here measured small numbers of photons. This brings to bear the possibility of measuring the QCS with other types of PNRDs, such as superconducting nanowire detectors that have recently been shown to discriminate between photon numbers up to four \cite{Divochiyetal2008,Mattiolietal2015,Zhuetal2020}. Such devices are faster than the transition-edge sensors used by Borealis, though they suffer from poorer confidence in determining photon numbers. If we perform the analysis on our data by artificially truncating the probability distributions at four photons, the resultant QCS values are always within 5\% of those reported in Table \ref{tab:res new}.  These imply that a variety of devices can be used to measure the QCS, which can be optimized for any given context.

Other interesting properties of quantum states can be measured using multiple copies of the state to be measured, configurable interferometers, and photon-number-resolving detectors, as detailed in Ref. \cite{Arnhemetal2022}. We measured $\expct{\had_{\mathrm{out}}\ha_{\mathrm{out}}}$ in the difference-mode at the output to estimate the overall transmission parameter $\eta$. This measurement is a direct measurement of  $\expct{\had_{\mathrm{in}}\ha_{\mathrm{in}}}-|\expct{\ha_{\mathrm{in}}}|^2$ for the input state \cite{Arnhemetal2022}, which indeed diminishes proportionally to the loss parameter. Moreover, if we were to measure $2\expct{\had_{\mathrm{out}}\ha_{\mathrm{out}}\otimes\hbd_{\mathrm{out}}\hb_{\mathrm{out}}}$ of the photon numbers at the two output ports, we would have measured $\expct{\had_{\mathrm{in}}^2\ha_{\mathrm{in}}^2}-|\expct{\ha_{\mathrm{in}}^2}|^2$ of the input, which would also diminish proportionally to the loss parameter. However, the latter quantity cannot be measured by assuming a constant loss throughout the circuit without loss of generality; moreover, neither constitutes a measure of quantumness. More copies of the state and more intricate circuits can be used to inspect a whole host of indicators of quantumness \cite{Arnhemetal2022} using machines like Borealis, which is well suited to these applications because it repeatedly produces identical quantum states in subsequent time bins that can then be made to interfere coherently with each other.

These devices can also be used to herald the production of other quantum states. For example, by heralding on measuring certain numbers of photons in one branch of the TMSV, one creates states such as have been studied in Ref. \cite{Thekkadathetal2020}. The QCS for such states was evaluated in Ref. \cite{Horoshkoetal2019}, which also require transmission parameters $\eta>1/2$ to certify quantumness, so this could be measured in another system with less overall loss to certify quantumness.

It is perhaps a coincidence that $\eta=1/2$ is the cutoff below which the QCS cannot certify quantumness for a variety of states. For initially pure Gaussian states with $\mathcal{C}^2>1$, one may prove (see Appendix \ref{app:lossy squeezed QCS}) that the QCS shrinks to unity exactly at $\eta=1/2$. One might expect this property to be true in general, as it holds true for non-Gaussian states such as Fock states and the aforementioned convex combinations of Fock states. However, it is not true for initially mixed Gaussian states. We thus may have the supposition that the QCS stops signifying quantumness when transmission reaches 50\% for all pure states and some mixed non-Gaussian states, but even that finds counterexamples in simple states like $(\ket{1}+\ket{2})/\sqrt{2}$. More study is certainly warranted of the sensitivity of this coherence measure to loss, as it may simultaneously constitute a transmission and a coherence quantifier.

\sugg{Is photon counting strictly necessary for measuring the QCS? Not if one has access to full state tomography, which can be done using heterodyne detection or by finding all of the moments of many different quadrature operators using homodyne detection. The problem with tomography is that it is expensive, slow, and error-prone: many different measurements are required for many different measurement settings in order to recreate a state's characteristic function, from which one can compute any property. Homodyne detection, for example, also requires a phase-stabilized local oscillator to be mode matched with the state being analyzed. In contrast, the setup here works efficiently for any input state and never requires changes in measurement settings; moreover, the detectors used here are highly efficient, with slight degradation due to multiplexing for the purpose of being sensitive to hundreds of modes \cite{Madsenetal2022}. It is also true that states guaranteed to be Gaussian only require measurements of the second-order moments in order to be fully characterized, so the strength of the current technique is that it is the only known one that works for arbitrary states and requires only a single measurement setting.}

At the heart of this \sugg{demonstration} is an array of PNRDs. The ability to reliably count and distinguish between different numbers of photons enables numerous quantum technologies, of which measuring the QCS is but one example. We expect both the QCS and devices like Borealis to find great application.

\acknowledgements{
The authors thank Stephan De Bi\`{e}vre, Duncan England, and Anaelle Hertz for helpful discussions.
The authors acknowledge that the NRC headquarters is located on the traditional unceded territory of the Algonquin Anishinaabe and Mohawk people. The authors acknowledge support from NRC's Quantum Sensors Challenge Program. This research was made possible through Innovative Solutions Canada's Testing Stream by providing access to Xanadu's Borealis device. AZG acknowledges funding from the NSERC PDF program.}

\sugg{All of the source code for deploying the circuits on Borealis and analyzing the data, all of the data files, and the device certificate, including details such as the squeezing strength, efficiencies of each delay loop and detector, static phases that are compensated for throughout the circuit, and more, are available on Github at \url{https://github.com/AaronGoldberg9/QCS_Borealis_public}.}

\appendix
\section{QCS for lossy Gaussian states}
\label{app:lossy squeezed QCS}
A single-mode Gaussian state has covariance matrix $\mathbf{V}$ with elements
\eq{
\begin{pmatrix}
    V_{xx}&V_{xp}\\ V_{xp}& V_{pp}
\end{pmatrix}\equiv\begin{pmatrix}
  \expct{\hat{x}^2}-\expct{\hat{x}}^2 &\expct{\frac{\hat{x}\hat{p}+\hat{p}\hat{x}}{2}}-\expct{\hat{x}}\expct{\hat{p}}\\
  \expct{\frac{\hat{x}\hat{p}+\hat{p}\hat{x}}{2}}-\expct{\hat{x}}\expct{\hat{p}} & \expct{\hat{p}^2}-\expct{\hat{p}}^2
\end{pmatrix}.
}  As an example, single-mode squeezed-vacuum states have 
$\expct{\hat{x}}=\expct{\hat{p}}=0$, with  \eq{
\mathbf{V}_{r\eu^{\iu\phi}}=\frac{1}{2}\cosh 2r\begin{pmatrix}
1&0\\0&1\end{pmatrix}-\frac{1}{2}\sinh 2r\begin{pmatrix}
\cos\phi&\sin\phi\\\sin\phi&-\cos\phi\end{pmatrix},
} while thermal states have diagonal covariance matrices with $\bar{n}+1/2$ for each variance on the diagonal.

Losing photons from such a state maintains its Gaussianity. This is represented by supplying another Gaussian state with a covariance matrix corresponding to the vacuum state, applying a beam splitter transformation to the joint covariance matrix, then ignoring the auxiliary state \cite{Brask2021arxiv}. The result is
\eq{
\mathbf{V}\to \mathbf{V}(\eta)=
\left(
\begin{array}{cc}
 \frac{1-\eta}{2}+V_{xx} \eta & V_{xp} \eta \\
 V_{xp} \eta & \frac{1-\eta}{2}+V_{pp} \eta \\
\end{array}
\right).
} This transformation and the property $\mathcal{P}=1/2\sqrt{\det \mathbf{V}}$ \cite{Parisetal2003} suffice for calculating the QCS of any lossy Gaussian state.

We use Eq. \eqref{eq:QCS Gaussian} to write, for any Gaussian state,
\eq{
\mathcal{C}^2(\hat{\rho}_{\mathrm{G}}; \eta)&=\frac{ \left[\eta (W-1)+1\right]\mathcal{P}_i^2}{\eta^2+(1-\eta) \left[\eta (2 W-1)+1\right]\mathcal{P}_i^2}\\
&=
1+\frac{\eta \left[\eta (1-2 \mathcal{P}_i^2 W+\mathcal{P}_i^2)+\mathcal{P}_i^2 (W-1)\right]}{\eta (2 (\eta-1) \mathcal{P}_i^2 W-\eta (\mathcal{P}_i^2+1)+2 \mathcal{P}_i^2)-\mathcal{P}_i^2},
\label{eq:QCS lossy Gaussian}
} where we have combined the total variance $W=V_{xx}+V_{pp}\geq 1$ and written the initial (i.e., $\eta=1$) purity as $\mathcal{P}_i$. We can inspect the zeroes of $\mathcal{C}^2(\hat{\rho}_{\mathrm{G}};\eta)-1$ to find when a quantum state subject to loss stops being certifiably quantum according to the QCS. The numerator reaches zero when either $\eta=0$, corresponding to the transmitted state being the vacuum, or when
\eq{
\eta^*=\frac{\mathcal{P}_i^2 (W-1)}{2 \mathcal{P}_i^2 W-\mathcal{P}_i^2-1}.
\label{eq:eta cutoff}
} This value is $1/2$ for initially pure states with $\mathcal{P}_i=1$ and grows monotonically with $\mathcal{P}_i$; when $W\mathcal{P}_i^2>1$ such that the initial state has $\mathcal{C}^2(\hat{\rho}_{\mathrm{G}})>1$, the state remains quantum according to the QCS for values of $\eta$ greater than $\eta^*$ given in Eq. \eqref{eq:eta cutoff}.

\end{document}